\PassOptionsToPackage{unicode}{hyperref}
\PassOptionsToPackage{hyphens}{url}
\PassOptionsToPackage{dvipsnames,svgnames,x11names}{xcolor}
\documentclass[
]{article}
\usepackage{amsmath,amssymb}
\usepackage{lmodern}
\usepackage{iftex}
\ifPDFTeX
  \usepackage[T1]{fontenc}
  \usepackage[utf8]{inputenc}
  \usepackage{textcomp} 
\else 
  \usepackage{unicode-math}
  \defaultfontfeatures{Scale=MatchLowercase}
  \defaultfontfeatures[\rmfamily]{Ligatures=TeX,Scale=1}
\fi
\IfFileExists{upquote.sty}{\usepackage{upquote}}{}
\IfFileExists{microtype.sty}{
  \usepackage[]{microtype}
  \UseMicrotypeSet[protrusion]{basicmath} 
}{}
\makeatletter
\@ifundefined{KOMAClassName}{
  \IfFileExists{parskip.sty}{%
    \usepackage{parskip}
  }{
    \setlength{\parindent}{0pt}
    \setlength{\parskip}{6pt plus 2pt minus 1pt}}
}{
  \KOMAoptions{parskip=half}}
\makeatother
\usepackage{xcolor}
\newlength{\cslhangindent}
\newlength{\csllabelwidth}
\newlength{\cslentryspacingunit} 
\newenvironment{CSLReferences}[2] 
 {
  \setlength{\parindent}{0pt}
  \ifodd #1
  \let\oldpar\par
  \def\par{\hangindent=\cslhangindent\oldpar}
  \fi
  \setlength{\parskip}{#2\cslentryspacingunit}
 }%
 {}
\usepackage{calc}

\ifLuaTeX
\usepackage[bidi=basic]{babel}
\else
\usepackage[bidi=default]{babel}
\fi
\babelprovide[main,import]{american}

\def\languageshorthands#1{}
\ifLuaTeX
  \usepackage{selnolig}  
\fi
\IfFileExists{bookmark.sty}{\usepackage{bookmark}}{\usepackage{hyperref}}
\IfFileExists{xurl.sty}{\usepackage{xurl}}{} 
\hypersetup{
  pdftitle={clustertools: A Python Package for Analyzing Star Cluster
Simulations},
  pdfauthor={Jeremy J. Webb},
  pdflang={en-US},
  colorlinks=true,
  linkcolor={Maroon},
  filecolor={Maroon},
  citecolor={Blue},
  urlcolor={Blue},
  pdfcreator={LaTeX via pandoc}}

\title{clustertools: A Python Package for Analyzing Star Cluster
Simulations}

\usepackage[affil-it]{authblk}
\author[1%
  ]{Jeremy J. Webb%
    }

\affil[1]{David A. Dunlap Department of Astronomy and Astrophysics,
University of Toronto, 50 St.~George Street, Toronto, ON, M5S 3H4,
Canada}
\date{6 May 2022}

\begin{document}
\maketitle

\hypertarget{summary}{%
\section{Summary}\label{summary}}

\texttt{clustertools} is a Python package for analyzing star cluster
simulations. The package is built around the \texttt{StarCluster} class,
which stores all data read in from the snapshot of a given model star
cluster. The package contains functions for loading data from commonly
used N-body codes, generic snapshots, and software for generating
initial conditions. All operations and functions within
\texttt{clustertools} are then designed to act on a
\texttt{StarCluster}. \texttt{clustertools} can be used for unit and
coordinate transformations, the calculation of key structural and
kinematic parameters, analysis of the cluster's orbit and tidal tails,
and measuring common cluster properties like its mass function, density
profile, and velocity dispersion profile (among others). While
originally designed with star clusters in mind, \texttt{clustertools}
can be used to study other types of \(N\)-body systems, including
stellar streams and dark matter sub-halos.

\hypertarget{statement-of-need}{%
\section{Statement of need}\label{statement-of-need}}

Stars do not form alone, but in clustered environments that in some
cases can remain gravitationally bound as a star cluster for billions of
years. The details of how exactly star clusters form, either at
high-redshifts or at present day, remain unknown. After formation, the
subsequent evolution of these stellar systems has been shown to be
strongly linked to that of their host galaxy. Hence star clusters are
often used as tools for studying star formation, galaxy evolution, and
galaxy structure. The comparison of simulated star clusters to
observations offers the ability to explore what formation conditions
reproduce the properties of observed star cluster populations and how
the evolution and structure of a galaxy affects star clusters.

A large number of \(N\)-body codes and software packages exist to
generate star cluster models and simulate their long-term evolution in
different environments (e.g.~AMUSE,
\protect\hyperlink{ref-amuse}{Portegies Zwart \& McMillan, 2018};
NBODY6, \protect\hyperlink{ref-nbody6}{Aarseth, 2003}; NBODY6++,
\protect\hyperlink{ref-nbody6pp}{Wang et al., 2015}; NEMO,
\protect\hyperlink{ref-nemo}{Teuben, 1995}; PETAR,
\protect\hyperlink{ref-petar}{Wang et al., 2020}). Additional software
exists for generating model star clusters directly from a known
distribution function (e.g.~GALPY, \protect\hyperlink{ref-galpy}{Bovy,
2015}; LIMEPY, \protect\hyperlink{ref-limepy}{Gieles \& Zocchi, 2015};
MCLUSTER, \protect\hyperlink{ref-mcluster}{Küpper et al., 2011}). The
output from these codes and packages can differ in units, coordinate
system, and format. The subsequent analysis of star cluster simulations
and models can then be quite inhomogeneous between studies, as various
methods exist for determining things like a cluster's centre, relaxation
time, or tidal radius. \texttt{clustertools} allows for data from a star
cluster simulation or model to be loaded and analyzed homogeneously by
first loading a \texttt{StarCluster} instance. Within a
\texttt{StarCluster}, stellar masses, positions, velocities,
identification numbers, and all relevant meta-data from the source is
saved. \texttt{clustertools} then contains an array of operations for
converting units and coordinate systems, functions for calculating key
cluster parameters, and functions for measuring how certain parameters
vary with clustercentric distance. In some cases, multiple different
methods are available for calculating a specific parameter. The software
is particularly useful to students or users new to star cluster studies,
as functions and profile measurements have a plotting feature that helps
illustrate how certain parameters are measured when possible. This
approach also ensures that the analysis of any star cluster is done in a
homogenous way with open-source code, regardless of the simulation or
model from which that data was produced.

\hypertarget{acknowledgements}{%
\section{Acknowledgements}\label{acknowledgements}}

JJW would like to thank Jo Bovy and Nathaniel Starkman for helpful
discussions and contributions to \texttt{clustertools}

\hypertarget{references}{%
\section*{References}\label{references}}
\addcontentsline{toc}{section}{References}

\hypertarget{refs}{}
\begin{CSLReferences}{1}{0}
\leavevmode\vadjust pre{\hypertarget{ref-nbody6}{}}%
Aarseth, S. J. (2003). \emph{{Gravitational N-Body Simulations}}.

\leavevmode\vadjust pre{\hypertarget{ref-galpy}{}}%
Bovy, J. (2015). {galpy: A python Library for Galactic Dynamics}.
\emph{Astrophysical Journal, Supplement}, \emph{216}(2), 29.
\url{https://doi.org/10.1088/0067-0049/216/2/29}

\leavevmode\vadjust pre{\hypertarget{ref-limepy}{}}%
Gieles, M., \& Zocchi, A. (2015). {A family of lowered isothermal
models}. \emph{Monthly Notices of the RAS}, \emph{454}(1), 576--592.
\url{https://doi.org/10.1093/mnras/stv1848}

\leavevmode\vadjust pre{\hypertarget{ref-mcluster}{}}%
Küpper, A. H. W., Maschberger, T., Kroupa, P., \& Baumgardt, H. (2011).
{Mass segregation and fractal substructure in young massive clusters -
I. The McLuster code and method calibration}. \emph{Monthly Notices of
the RAS}, \emph{417}(3), 2300--2317.
\url{https://doi.org/10.1111/j.1365-2966.2011.19412.x}

\leavevmode\vadjust pre{\hypertarget{ref-amuse}{}}%
Portegies Zwart, S., \& McMillan, S. (2018). \emph{{Astrophysical
Recipes; The art of AMUSE}}.
\url{https://doi.org/10.1088/978-0-7503-1320-9}

\leavevmode\vadjust pre{\hypertarget{ref-nemo}{}}%
Teuben, P. (1995). {The Stellar Dynamics Toolbox NEMO}. In R. A. Shaw,
H. E. Payne, \& J. J. E. Hayes (Eds.), \emph{Astronomical data analysis
software and systems IV} (Vol. 77, p. 398).

\leavevmode\vadjust pre{\hypertarget{ref-petar}{}}%
Wang, L., Iwasawa, M., Nitadori, K., \& Makino, J. (2020). {PETAR: a
high-performance N-body code for modelling massive collisional stellar
systems}. \emph{Monthly Notices of the RAS}, \emph{497}(1), 536--555.
\url{https://doi.org/10.1093/mnras/staa1915}

\leavevmode\vadjust pre{\hypertarget{ref-nbody6pp}{}}%
Wang, L., Spurzem, R., Aarseth, S., Nitadori, K., Berczik, P.,
Kouwenhoven, M. B. N., \& Naab, T. (2015). {NBODY6++GPU: ready for the
gravitational million-body problem}. \emph{Monthly Notices of the RAS},
\emph{450}(4), 4070--4080. \url{https://doi.org/10.1093/mnras/stv817}

\end{CSLReferences}

\end{document}